\documentclass[
    ,final            
  ]
  {aipproc}

\layoutstyle{8x11double}
\usepackage{array}
\usepackage{color}
\usepackage{mdframed}

\definecolor{red}{rgb}{1,0,0}
\definecolor{orange}{rgb}{1,0.5,0}
\definecolor{green}{rgb}{0.13,0.55,0.13}
\definecolor{purple}{rgb}{0.5,0,1}
\definecolor{mag}{rgb}{1.0,0.0,1.}

\begin{document}

\title{Peer Evaluation of Video Lab Reports in a Blended Introductory Physics Course}


\author{Scott S. Douglas}{
  address={Department of Physics, Georgia Institute of Technology, Atlanta, GA 30332}
  }

\author{Shih-Yin Lin}{
  address={Department of Physics, National Changhua University of Education, Changhua, Taiwan 500}
  }

\author{John M. Aiken}{
  address={Department of Physics, Georgia Institute of Technology, Atlanta, GA 30332}
  }

\author{Brian D. Thoms}{
  address={Department of Physics and Astronomy, Georgia State University, Atlanta, GA 30303}
  }

\author{Edwin F. Greco}{
  address={Department of Physics, Georgia Institute of Technology, Atlanta, GA 30332}
  }

\author{Marcos D. Caballero}{
  address={Department of Physics and Astronomy, Michigan State University, East Lansing, MI 48824}
  }

\author{Michael F. Schatz}{
  address={Department of Physics, Georgia Institute of Technology, Atlanta, GA 30332}
  }

\begin{abstract}
The Georgia Tech blended introductory calculus-based mechanics course emphasizes scientific communication as one of its learning goals, and to that end, we gave our students a series of four peer-evaluation assignments intended to develop their abilities to present and evaluate scientific arguments. Within these assignments, we also assessed students' evaluation abilities by comparing their evaluations to a set of expert evaluations. We summarize our development efforts and describe the changes we observed in student evaluation behavior.
\end{abstract}

\maketitle

\section{\label{sec:intro}Introduction}

In Spring 2014, Georgia Tech took two sections of its large-enrollment introductory mechanics course using the Matter \& Interactions (M\&I) curriculum \cite{MandI} and ran a ``blended'' version of that same course (N=355 students). Our blended course featured out-of-classroom laboratory exercises with online video lab reports and peer evaluation. Traditional lectures were largely replaced with online lecture videos \cite{YWYLyoutube}, but unlike a fully ``flipped'' course, our blended course still devoted some time to formal instructor lecturing.

We intended our course to serve two important learning goals: to help our students develop an understanding of physics as something applicable to their everyday experience, and to develop their practice of scientific communication (both in presenting their own findings and evaluating their peers' presentations). To serve the first goal, we designed four laboratory activities with an eye toward computational modeling and real-world physics practice. To serve the second goal, we instructed our students to produce video lab reports in the style of a short colloquium talk and participate in an anonymous peer evaluation process.

This paper describes our progress toward the latter part of this second learning goal. How does our students' peer evaluation behavior change over the course of the semester? To address this question, we compared student and instructor ratings of 20 video lab reports across four labs. Our proximate goal (and the goal of this paper) is to characterize this change numerically, and to deal with potential sources of systemic bias.

\vspace*{-14pt}
\section{\label{sec:labcycle}Our Labs}

In a typical lab, students were instructed to perform individually an authentic observational activity inspired by the M\&I curriculum. Students then used the software tools Tracker \cite{tracker} and VPython \cite{vpython} to analyze and model their observational systems computationally. Finally, each student prepared a five-minute video lab report in the style of a short colloquium talk. Students had two weeks to perform each laboratory activity, prepare their lab reports, and upload their lab reports to YouTube and submit the link to the course instructors. 


In the week following the submission of the lab reports, we conducted the peer evaluation process with the rubric summarized in Table \ref{table:rubric} (students were shown this rubric in-class before beginning their first lab assignment). The rubric asks students to evaluate each video lab report in terms of its structure, its physics content, and its production quality. Each of five items on the rubric comprised one rating on a five-point poor-to-excellent scale and one textual comment. Students received significant support and instruction before doing their first evaluations; our corpus of lecture videos contained videos about preparing and evaluating video lab reports, including a step-by-step example evaluation of two actual video lab reports and several videos specifically relevant to each laboratory exercise \cite{YWYLyoutube}. In the classroom, we supplemented these instructional videos with in-class small-group practice presentations intended to provide students with helpful feedback on their lab reports in progress.

\begin{table}
\caption{Rubric Summary for Spring 2014}
\centering
\begin{tabular}{>{\centering\arraybackslash}m{0.1cm} >{\centering\arraybackslash}m{1.4cm} >{\arraybackslash}m{5.1cm}}
\hline\hline
{\bf Item} & {\bf Topic} & {\bf Features to Evaluate} \\ [0.5ex] 
\hline
1& Organization/ Structure&Introduction, conclusion, transitions, overall flow of presentation \\ \hline
2& Content: Models&Identification of relevant models, discussion of main physics concepts, connection between models \& fundamental principles \\ \hline
3& Content: Prediction Discussion & Comparison between computer-generated data and observational data, explanation of discrepancies between same \\ \hline
4& Content: Overall & Presence or absence of major physics errors, discussion of specific instructor-posed questions\\ \hline
5& Production/ Delivery & Technical quality of video, vocal quality of narration \\[1ex]
\hline
\end{tabular}
\label{table:rubric}
\end{table}

As part of the peer evaluation process, course instructors randomly assigned three peer videos to each student, along with five common instructor-evaluated videos and each student's own video (for self-evaluation) for a total of nine video evaluations per student per lab. Evaluations were conducted in three phases:

\begin{enumerate}
	\item {\bf Practice:} Each student evaluated two of the common instructor-evaluated videos for practice (no credit). After evaluating each of the two practice videos (hereafter {\bf P1} \& {\bf P2}), each student was shown the detailed instructor evaluation for that video.
	\item {\bf Calibration:} Each student evaluated another two of the common instructor-evaluated videos ({\bf C1} \& {\bf C2}) for credit, and received a ``calibration grade'' dependent on how well her evaluation aligned with the expert evaluation for that video. After evaluating each calibration video, each student was shown the detailed instructor evaluation for that video.
	\item {\bf Evaluation:} Each student evaluated three peer videos, her own video, and a final instructor-evaluated video in random order. This last instructor-evaluated video (the ``hidden'' calibration video, {\bf HC}) was presented to the student as just another peer video, but her evaluation of this video also counted toward her calibration grade. Students were never shown the instructor evaluation for the hidden calibration video.
\end{enumerate}

\begin{figure*}
\includegraphics{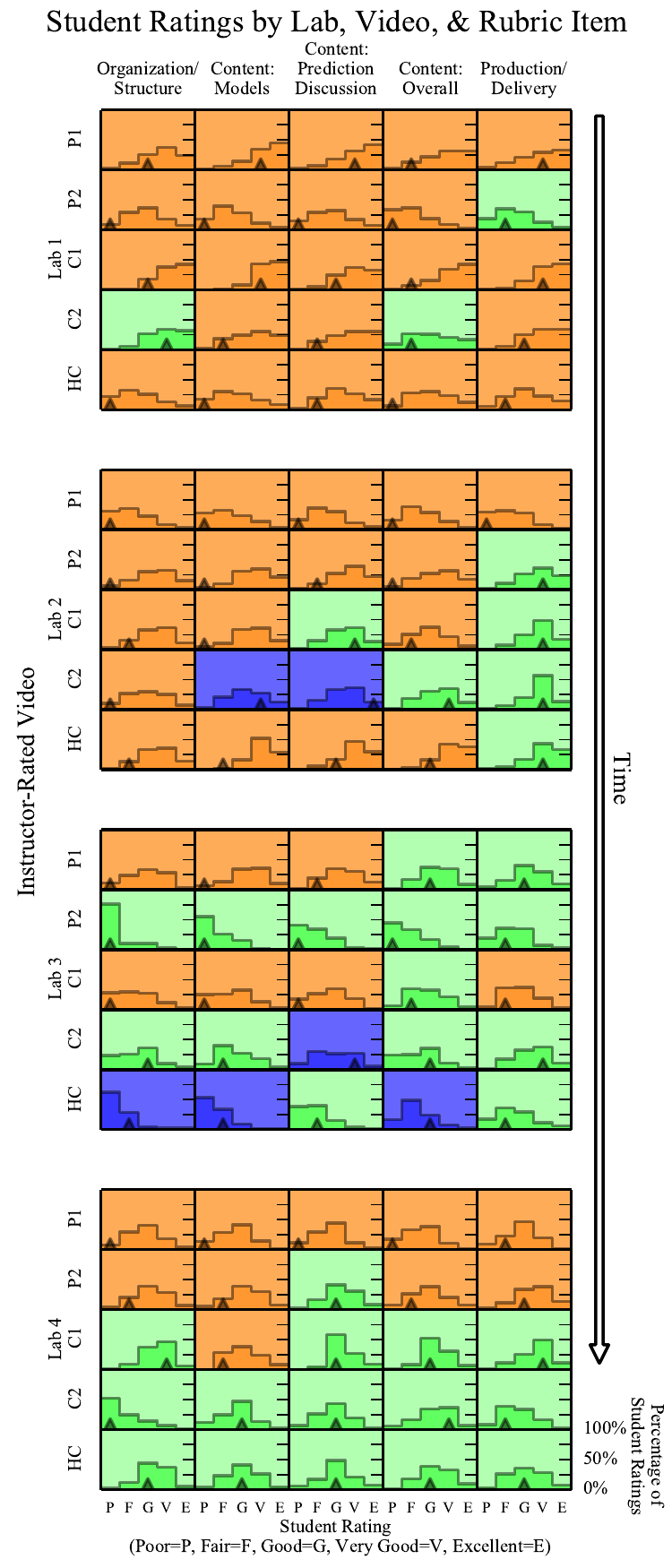}
\caption{Student ratings for all instructor-rated videos in Spring 2014. Each column represents one rubric item as described in Table \ref{table:rubric}, each row represents one video, and each cell shows the normalized distribution of student ratings for that item on that video. Green (light) cells indicate instances where the modal student rating was the same as the instructor rating; blue (dark) and orange (medium) cells indicate where the modal student rated below or above the instructors, respectively. Instructor ratings are indicated with dark triangles. Reading the figure from top to bottom shows more instructor-student agreement as the semester progresses.}\label{fig:ratings}
\end{figure*}

\section{\label{sec:results}Results}

304 students completed all four lab assignments. Figure \ref{fig:ratings} shows the distribution of student ratings for every rubric item for every instructor-rated video in the semester. Reading the figure from top to bottom follows the students' chronological progress through all the instructor-rated videos in the course; read this way, Fig. \ref{fig:ratings} shows an apparent overall gain in agreement between student and expert ratings on instructor-rated videos over the semester.  The increasing proportion of green cells toward the bottom of Fig. \ref{fig:ratings} represents an increasing number of instances where a plurality of students agreed with the instructor rating for a rubric item on a video. However, the instructor ratings themselves were not equally distributed among all the labs; the videos in Lab 4, for example, had more ``good'' instructor ratings (N=11) than did the videos in Lab 1 (N=4). If students were biased toward responding with e.g. ``good'', then the apparent gain in agreement may have been an artifact of a systemic bias caused by the chance presence of more ``good'' instructor ratings toward the end of the course.

\begin{figure*}

\includegraphics{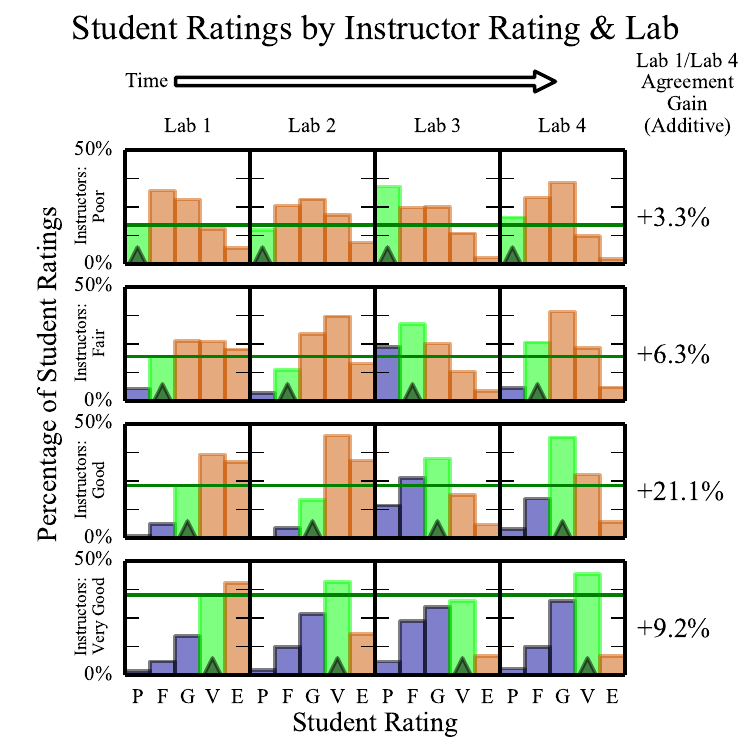}

\caption{Student ratings for all instructor-rated videos in Spring 2014, sorted by instructor rating. A horizontal line indicates the proportion of agreement in Lab 1. Each row represents one instructor rating, each column represents one lab, and each cell shows the normalized distribution of student ratings for all items in that lab which received that instructor rating. Instructor ratings are indicated with dark triangles. Green (light) bars represent the proportion of students who gave the same rating as the instructors within each cell, while blue (dark) and orange (medium) represent the proportion of students who rated below or above the instructors, respectively. Comparing the rightmost column to the leftmost column shows at least a small gain in agreement across all instructor ratings, with the largest gain in ``good''.}\label{fig:byscore}

\end{figure*}

Figure \ref{fig:byscore} addresses this concern by looking at student ratings between labs but within expert ratings, excluding the single ``excellent'' among all 100 instructor ratings. We excluded the instructor rating ``excellent'' because of concerns over statistical power and because a single instructor rating in one lab does not allow us to do any cross-lab comparisons within that rating. The differences between all student rating distributions in Lab 1 and subsequent labs were determined to be statistically significant with a Kolmogorov-Smirnov two-sample test, p << 0.05 (the K-S two-sample test is a nonparametric test suitable for comparing distributions of ordinal data) \cite{KS2sample}. Read from left to right, this figure shows at least a small gain in student/instructor agreement within every instructor rating and a large gain within ``good'', precluding the possibility that the overall gain in agreement is due solely to a systemic bias caused by a nonuniform distribution of instructor ratings. The mean student-instructor agreement within ``poor'' (mean 22\%, range 15\%-34\%) and ``fair'' (mean 23\%, range 14\%-34\%) is substantially less than the mean student-instructor agreement within ``very good'' (mean 38\%, range 32\%-44\%), and is indeed barely above chance (20\% agreement for uniform random guessing). As it turns out, the presence of more ``very good'' responses in Lab 1 than in Lab 4 actually serves to make the apparent gain artificially low. Student-instructor agreement within ``good'' (mean 30\%, range 23\%-44\%) improves from Lab 1 to Lab 4 twice as much as within any other rating, has in Lab 4 the same proportion of agreement as does ``very good'' (44\%), and is the only instructor rating which shows a >10\% gain in instructor-student agreement.

Overall, the data exhibit three major trends. Most of the gain in agreement occurred in the middle of the rating scale (i.e., when the instructors and students said ``good''), students almost stopped giving ``excellents'' toward the end of the semester, and agreement at the high end of the scale (``very good'') began high and stayed high while agreement at the low end (``poor'' and ``fair'') began low and stayed low.

\section{\label{sec:discussion}Discussion}

Our analysis so far has characterized the change in student ratings without addressing the reasons behind those changes. We posit two possible causes which may be responsible for the observed changes in student ratings.

Firstly, we posit students may be adopting instructors' norms regarding the values on the rating scale. The three trends described in the Results section suggest that the overall gain in agreement may be partly explained by a change in student understanding of the rating scale itself (e.g., how much better than ``very good'' is ``excellent''?). There was only one ``excellent'' among all 100 instructor ratings in the course. This reflects an instructor norm that holds ``excellent'' to be a much loftier rating than our students might have initially thought, given that ``excellent'' constituted 24\% of student ratings in Lab 1. If our students had learned to avoid giving ``excellent'' ratings, then this trend alone may account for some of the overall gain in student-instructor agreement. As illustrated by both Figs. \ref{fig:ratings} and \ref{fig:byscore}, the ``excellents'' among the student ratings indeed almost disappear after Lab 2, even though the instructor ratings for the latter two labs are not lower overall, suggesting that our students might have adopted a more instructor-like norm regarding ``excellent''.

Secondly, we posit that students may be learning to attend to different features within the videos in a more instructor-like manner, thereby adopting instructors' video-watching practices. While the overall trend shown in Fig. \ref{fig:ratings} is toward increased agreement, there are still some remarkable instances of disagreement between student and instructor ratings throughout the semester which should shed some light on why students and instructors give the ratings they do. For example, Lab 3 Practice 1 Item 1 (L3P1\#1) has an instructor rating of ``poor'' but shows low student-instructor agreement and low student-student agreement. L3P2\#1, the same item on the very next video, also has an instructor rating of ``poor'' but shows a very different distribution of student responses; student-instructor and student-student agreement is very high, since almost all students also rated this item ``poor''. L2P2\#3 and Lab 2 Calibration 2 Item 3 (L2C2\#3) show the opposite phenomenon on a different rubric item. Here, the distributions of student ratings are roughly similar, but the instructor ratings are very different. These ratings suggest a difference in the actual content of those videos and a difference between the video features to which instructors and students attend when evaluating that rubric item (i.e., our posited difference in video-watching practices). This is corroborated by instructor comments; instructors rated L3P1\#1 ``poor'' because the introduction was {\it ``simply [a] reading [of] the problem statement''}, but gave L3P2\#1 the same ``poor'' rating for a different reason (L3P2 contained {\it ``no intro at all''}).

We cannot fully explore students' video-watching practices or confirm any changes in student rating-scale norms by analyzing the ratings alone. To get a clear picture of our students' rating norms and video-watching practices, we will need to examine students' comments, conduct student interviews, and code the content of the video lab reports. We expect student comments to provide a rich source of information for our ongoing work; these comments span the range from terse declaratory statements to detailed explanations of the reasons behind the ratings. For example, a majority of students wrote some variation of {\it ``no introduction''} on L3P2\#1, while one student commented on L3P2\#2 {\it ``Pretty good here! I put good instead of very good because my physics professor told me to be mean when grading.''} In the process of coding features within the video lab reports, we will also address one more potential source of systemic bias: uneven distribution of different video features throughout the semester. When we look within instructor ratings (as we do in Fig. \ref{fig:byscore}), we are not necessarily looking within specific video features, since the same instructor rating can be associated with very different instructor comments.

\section{\label{sec:conclusion}Concluding Remarks}

The gain in student/instructor agreement in these video evaluations is real in that it is not solely a result of a systemic bias introduced by the chance distribution of instructor ratings among videos throughout the four labs. Our students got at least slightly better at agreeing with instructor ratings across the board, but got substantially higher agreement only among the items which instructors rated ``good''. The instructor ratings of ``poor'' and ``fair'' exhibited lower overall instructor-student agreement than did ``very good''. We do not yet have a clear picture of what norms and practices related to peer evaluation our students are adopting to produce these gains, because the identification of the specific norms and practices that inform peer evaluation lie beyond any analysis of the ratings alone.

In our future work, we will attempt to characterize this gain in rating agreement in terms of the norms and practices adopted by our students throughout the semester. We have already begun an investigation of the comments left by students along with each of their ratings, which we expect will shed some light on how and why the reasoning, practices, and norms behind the student ratings evolve during the course of instruction. 

This work was supported by the Gates Foundation and the Georgia Governor's Office of Student Achievement. We gratefully acknowledge the work of Christopher Wang and other members of the Georgia Tech Physics MOOC VIP team who helped develop the peer evaluation software we used in our course. We further acknowledge the work of Dr. David Lawrence and the Georgia Tech Center for the Enhancement of Teaching and Learning in helping to develop our rubric.



\bibliographystyle{aipproc}   

\bibliography{ccmi}

\IfFileExists{\jobname.bbl}{}
 {\typeout{}
  \typeout{******************************************}
  \typeout{** Please run "bibtex \jobname" to optain}
  \typeout{** the bibliography and then re-run LaTeX}
  \typeout{** twice to fix the references!}
  \typeout{******************************************}
  \typeout{}
 }

\end{document}